\documentclass[preprint]{aastex631}

\usepackage{hyperref}
\usepackage{xspace}

\newcommand{\kepler}{\emph{Kepler}\xspace}
\newcommand{\ktwo}{\emph{K2}\xspace}

\shorttitle{Enabling Exoplanet Demographics Studies}
\shortauthors{SIG\#2: Exoplanet Demographics}
\graphicspath{{./}{figures/}}
\usepackage{soul} % Strikethrough with \st{}
\definecolor{gijs}{RGB}{31,120,180}

\definecolor{eliot}{RGB}{200,60,122}

\newcommand{\mathbold}[1]{\mbox{\boldmath $\bf#1$}}
\newcommand\piEbold{{\mathbold \pi_{\rm E}}}

\newcommand\murelbold{{\mathbold \mu_{\rm rel}}}

\begin{document}

\title{Enabling Exoplanet Demographics Studies with Standardized Exoplanet Survey Meta-Data}
%\title{SIG 2: Findings on the Importance of Exoplanet Survey Meta-Data}
%\title{SIG 2: The Urgent Need for Standardized Exoplanet Survey Meta-Data}
%\title{SIG 2: Addressing the Need for Standardized Exoplanet Survey Meta-Data}

\correspondingauthor{Jessie Christiansen}
\email{christia@ipac.caltech.edu}

\author{Prepared by the ExoPAG Science Interest Group (SIG) \#2 on Exoplanet Demographics}
\noaffiliation

\author{Jessie L. Christiansen}
\affiliation{Caltech/IPAC-NASA Exoplanet Science Institute, Pasadena, CA 91125, USA}

\author{David P. Bennett}
\affiliation{Laboratory for Exoplanets and Stellar Astrophysics, NASA Goddard Space Flight Center,
Greenbelt, MD 20771, USA}
\affiliation{Department of Astronomy, University of Maryland, College Park, MD 20742, USA}

\author{Alan P. Boss}
\affiliation{Earth \& Planets Laboratory, Carnegie Institution for Science, 5241 Broad Branch Road,
NW, Washington, DC 20015-1305}

\author{Steve Bryson}
\affiliation{NASA Ames Research Center, Moffett Field, CA 94035, USA}

\author{Jennifer A. Burt}
\affiliation{Jet Propulsion Laboratory, California Institute of Technology, 4800 Oak Grove Drive, Pasadena, CA 91109, USA}

\author{Rachel B. Fernandes}
\affiliation{Lunar and Planetary Laboratory, The University of Arizona, Tucson, AZ 85721, USA}

\author{Todd J. Henry}
\affiliation{RECONS Institute, Chambersburg, PA 17201, USA}

\author{Wei-Chun Jao}
\affiliation{Department of Physics and Astronomy, Georgia State University, Atlanta, GA 30303, USA}

\author{Samson A. Johnson}
\affiliation{Department of Astronomy, The Ohio State University, 140 West 18th Avenue, Columbus OH 43210, USA}
\affiliation{Jet Propulsion Laboratory, California Institute of Technology, 4800 Oak Grove Drive, Pasadena, CA 91109, USA}

\author{Michael R. Meyer}
\affiliation{Department of Astronomy, University of Michigan, Ann Arbor, MI USA 48109}

\author{Gijs D. Mulders}
\affiliation{Facultad de Ingenier{\'i}a y Ciencias, Universidad Adolfo Ib{\'a}{\~n}ez, Av. Diagonal las Torres 2640, Pe{\~n}alol{\'e}n, Santiago, Chile}
\affiliation{Millennium Institute for Astrophysics, Chile}

\author{Susan E. Mullally}
\affiliation{Space Telescope Science Institute, 3700 San Martin Drive, Baltimore, MD 21218, USA}

\author{Eric L. Nielsen}
\affiliation{Kavli Institute for Particle Astrophysics and Cosmology, Stanford University, Stanford, CA 94305, USA}

\author{Ilaria Pascucci}
\affiliation{Lunar and Planetary Laboratory, The University of Arizona, Tucson, AZ 85721, USA}

\author{Joshua Pepper}
\affiliation{Department of Physics, Lehigh University, 16 Memorial Drive East, Bethlehem, PA 18015, USA}

\author{Peter Plavchan}
\affiliation{Department of Physics and Astronomy, George Mason University, Fairfax, VA 22030, USA}

\author{Darin Ragozzine}
\affiliation{Department of Physics \& Astronomy, N283 ESC, Brigham Young University, Provo, UT 84602, USA}

\author{Lee J. Rosenthal}
\affiliation{MIT Lincoln Laboratory, 244 Wood Street, Lexington, MA 02421-6426}

\author{Eliot Halley Vrijmoet}
\affiliation{Department of Physics and Astronomy, Georgia State University, Atlanta, GA 30303, USA}
\affiliation{RECONS Institute, Chambersburg, PA 17201, USA}

\section{Executive Summary}
\label{sec:summary}

Goal 1 of the National Academies of Science, Engineering and Mathematics Exoplanet Science Strategy\footnote{\url{https://www.nap.edu/catalog/25187/exoplanet-science-strategy}}, which was endorsed by the Decadal Survey on Astronomy and Astrophysics 2020, is \emph{``to understand the formation and evolution of planetary systems as products of the process of star formation, and characterize and explain the diversity of planetary system architectures, planetary compositions, and planetary environments produced by these processes"}, with the finding that \emph{``Current knowledge of the demographics and characteristics of planets and their systems is substantially incomplete."} One significant roadblock to our ongoing efforts to improve our demographics analyses is the lack of comprehensive meta-data accompanying published exoplanet surveys. The Exoplanet Program Analysis Group (ExoPAG) Science Interest Group \#2: Exoplanet Demographics has prepared this document to provide guidance to survey architects, authors, referees and funding agencies as to the most valuable such data products for five different exoplanet detection techniques--transit, radial velocity, direct imaging, microlensing and astrometry. We find that making these additional data easily available would greatly enhance the community’s ability to perform robust, reproducible demographics analyses, and make progress on achieving the most important goals identified by the exoplanet and wider astronomical community.

\section{Introduction}
\label{sec:intro}

The field of exoplanets has now firmly moved from detection and characterization of individual objects and systems to characterization of entire populations of objects and systems. By studying the demographics of exoplanetary systems, we learn more about the physical mechanisms that drive their formation \citep[e.g.][]{Mordasini2016,Lee2016,Jin2018}, migration \citep[e.g.][]{Dawson2013,Yu2021}, and evolution \citep[e.g.][]{Owen2017,Ginzburg2018}. 

The exoplanet detection techniques that have provided sufficient detections for population studies cover very different areas of parameter space in both stellar and planet properties. Observational biases push the different detection techniques toward different populations of stellar hosts (across properties such as distance, mass, age, activity, stellar multiplicity), and different populations of planets (across such properties as radius, mass, orbital period, planet multiplicity, and proximity to resonance). For example, transit and radial velocity surveys are most sensitive to close-in planets ($<$1\,au), microlensing surveys and radial velocity to planets around the ice line ($\sim$1--10\,au), and direct imaging surveys to distant planets ($>$10\,au).

Although the detection techniques probe different pieces of parameter space, together they produce a broad census of planet populations. Exploiting this coverage to produce a comprehensive understanding of planet demographics is a crucial input to the planning and design of future NASA mission prioritized by the 2020 Decadal Survey on Astronomy and Astrophysics\footnote{\url{https://www.nationalacademies.org/our-work/decadal-survey-on-astronomy-and-astrophysics-2020-astro2020}}, and to achieving NASA's strategic objectives of ``discovering the secrets of the universe'' and ``searching for life elsewhere''\footnote{\url{https://science.nasa.gov/science-red/s3fs-public/atoms/files/nasa_2018_strategic_plan_0.pdf}}. The challenge is to stitch these different pieces together to provide a robust, overarching understanding of the demographics of planets and planetary systems. Such a process requires disentangling the stellar and planet population biases inherent to each technique, as well as correcting for the detection sensitivity and reliability of each individual survey. In addition, survey results are often reported in terms of planet occurrence rates, but these rates depend on assumptions made about the underlying planet population, and these assumptions are not always consistent across detection methods and surveys, leading to diverging results. 

There have been several notable attempts to combine demographics results from multiple detection techniques \citep[e.g.][]{Clanton2014,Clanton2016,KunimotoBryson2021}.
%, confined thus far to M dwarf host stars. [commented out by Gijs]
Other studies have compared the results of different detection techniques without attempting to fit them with a single planet population. Examples include Fig.\ 2 of \cite{pascucci2018universal}, which compares transit and microlensing results, Fig.\ 2 of \citet{Fernandes2019}, which compares transit and radial velocity results, Fig.\ 8 of \citet{Nielsen2019}, which compares radial velocity and direct imaging results, and Figures 9 and 10 of \cite{2021ApJS..255...14F} which compare radial velocity vs. direct imaging and vs. microlensing. As yet, there have been few wide-ranging analyses that take advantage of the survey data that are currently available.

The difficulty in moving forward with synthesizing results from multiple detection techniques is typically that these survey data are not accompanied with the meta-data necessary to be reliably incorporated with other surveys. This includes, for instance, information about the surveys themselves, the stars observed, and the criteria for planets detected. There are certainly hurdles to making these data available: identifying the types and formats of the ancillary information needed; performing the work to create the data products, and making them readily accessible once published. However, there is significant benefit to undertaking this effort, including a substantial increase in the use, visibility, and citations of the accompanying survey data. For example, the NASA \kepler mission provided a wealth of additional information characterizing the survey\footnote{\href{https://exoplanetarchive.ipac.caltech.edu/docs/Kepler_completeness_reliability.html}{NASA Exoplanet Archive: Kepler Completeness and Reliability Products}} \citep[e.g][]{Thompson2018,Christiansen2020}, allowing many teams across the community to perform their own independent investigations of the data \citep[e.g.][]{Fernandes2019,Mulders2015,Mulders2018,Hsu2019,Zink2019b}. 

Another challenge with missing meta-data arises when moving beyond demographics of individual planets to understanding the demographics of planetary system architectures, where correlations between planets within a particular system are of high scientific value but can often be biased by the survey or analysis technique \citep[e.g.][]{Zink2021}. As we continue to move into a regime where planetary systems have different planets detected by different techniques, improved understanding of survey properties with respect to multi-planet systems will be crucial. 

{\bf The goal of this document is to lay out, in some detail, the necessary ancillary information for five different exoplanet detection techniques---transit, radial velocity, direct imaging, microlensing, and astrometry---that could be published alongside exoplanet survey data to ensure those data are maximally useful to the demographics community.} In particular, we find that intermediate data products generated as part of occurrence rate calculations are often not published, but are instrumental in combining data both from surveys using the same detection technique, and across multiple detection techniques. As such, we find that making these additional data easily available would greatly enhance the community's ability to perform robust, reproducible occurrence rate calculations.

The audience for this document includes, but is not limited to:
\begin{itemize}
    \item Authors publishing exoplanet survey data;
    \item Referees for journals publishing exoplanet survey data;
    \item Funding agencies considering proposals that would fund the collection, analysis, and publication of exoplanet survey data
\end{itemize}

This document was created by the NASA Exoplanet Program Analysis Group (ExoPAG) Science Interest Group (SIG) on Exoplanet Demographics. It includes input from astronomers publishing exoplanet survey data across all the major detection techniques, and those actively working in exoplanet demographics. In assembling these recommendations, we identified two `tiers' of data products. We find that availability of {\bf Tier I} data products would enable rudimentary incorporation of published exoplanet survey data into other analyses, using either the same or different detection techniques. Tier I products are those which are typically readily at hand, and are required for someone to be able to make some attempt to integrate the data in their analysis. We find that availability of {\bf Tier II} data products would greatly enhance the community’s ability to incorporate published exoplanet survey data into their analyses, using either the same or different detection techniques. Tier II data products are those which may require more effort to produce or archive due to computational and data volume limitations, but would enable a robust inclusion and maximum usefulness of the published data set. In addition, for each detection technique we highlight in bold the data product that would be most useful for incorporating the survey results in demographics analyses, in order to help the community prioritize their resources.

An illustrative example is the survey detection completeness: while the survey-averaged completeness as described in the Tier I data products allows for a reproduction of survey results or a re-analysis with a different methodology, the per-star completeness described in the Tier II data products allows for new analyses of specific sub-samples of stars within the survey. The latter provides the community much greater flexibility in, for instance, combining targets from different surveys and accounting for target overlaps between surveys. In addition to the tiers, we identify three data `types', for ease of categorizing: data pertaining to the {\bf stellar sample}, data pertaining to the {\bf survey properties}, and data pertaining to the {\bf planet catalog}.

One driving philosophy in defining the products below is the goal that they are kept as close to the `observable' units as possible. For instance, each detection technique measures a different property of the exoplanet size: transits measure the planet radius (or, more correctly, the planet-to-star radius ratio); radial velocity measures the planet mass projected along the line of sight with respect to the stellar mass; microlensing and astrometry measure the planet-to-star mass ratio; and direct imaging measures the planet luminosity in a given waveband. Each of these planet properties can be mapped onto the others through a variety of models and relations, however different models and relations may give different results and some of these relations are probabilistic in nature, \citep[e.g.][]{2016ApJ...825...19W}, and groups will have preferences as to the models and relations they wish to employ. We find that publication of survey and accompanying data in these native units would allow for future refinement of models and relations as our understanding improves. In addition, where authors have transformed their observables into other properties, we find that supply of the models and/or assumptions used in the transformation allows for a more extensive utility of the data set. Keeping the data in their native observable units also more naturally allows for updates as stellar parameters, such as the age, distance, or mass, are updated.

A second driving philosophy is that of reproducibility: survey data are often published without sufficient accompanying data for their demographics results to be reproduced by independent teams. Part of the motivation for this document is to enable authors to provide the data required for the demographics community to reproduce their results. In that vein, we find that the public availability of analysis codes used in publications greatly improves their reproducibility. Finally, we find that providing the described data products in machine-readable format (as compared to figures from which the relevant data would need to be extracted) is also crucial for allowing reproducibility. 

For each detection technique below, we provide below the Tier I and Tier II data products for the three data types. We find that there is significant new exoplanet science to be explored if these data are more readily available. We also invite feedback on these lists, particularly if there is anything we have overlooked.

%\section{Transit surveys}
%\label{sec:transit}

\section[Transits]{Transits%
              \sectionmark{Transits}}
\sectionmark{Transits}

The transit method measures the ratio of the planetary radius to the stellar radius, as well as the orbital period of exoplanets. Because the transit probability for an individual system is low, transit surveys are typically designed to monitor large number of stars in order to detect exoplanets. The detection probability of a transiting planet is a strong function of its orbital period and planet radius, ranging from $\sim$10\% for hot Jupiters to below a fraction of a percent for (super-)Earth-sized planets in the habitable zones around sun-like stars \cite[see, e.g.][]{Borucki1984}.

To do population studies, that measure either the intrinsic occurrence of planets or the dependence of occurrence on planet properties, the planet detection efficiency---the fraction of planets at a given detection threshold successfully recovered---has to be estimated. This can be done approximately by using a transit signal-to-noise threshold, above which one assumes 100\% detection efficiency \citep[e.g.][]{Howard2012}, or empirically by using injection-recovery tests to measure real pipeline performance \citep[e.g.][]{Thompson2018,Christiansen2020}.

Ground-based and space-based transit surveys employ different observation strategies and probe different parts of exoplanet parameter space. Space-based observatories allow for continuous monitoring of large groups of stars at high sensitivity. For example, the \kepler spacecraft monitored almost 200,000 stars for transiting planets as small as Earth with periods up to a year, enabling precise estimates of the occurrence of planets and fraction of stars with planetary systems within 1\,au \citep[e.g.][]{Mulders2018,Hsu2019,He2019}. A similar dataset with over $200,000$ stars from the \ktwo mission is also available \citep{Zink2021}. The Transiting Exoplanet Survey Satellite (TESS, \citealt{Ricker2015}) is significantly increasing the number of stars with continuous, high-precision time series photometry with millions of light curves from an all sky survey \citep{Kunimoto2021LC} and identifying thousands of transiting planet candidates \citep[e.g.][]{Kunimoto2021PC}.

Ground-based observatories are limited in the time and number of stars they can monitor. They have been very efficient in discovering hot Jupiters, % is there a HJ survey overview paper to cite here?
or targeting specific sub-sets of stars such as M dwarfs to look for habitable zone planets \citep{Nutzman2008,Jehin2011}, but have been less proficient in providing uniform catalogues for planet occurrence rate studies.

%Request from microlensing surveys: 1. Mass-radius relations for cooler planets with low stellar irradiation.

Given the large number of stars in transit surveys, the survey detection efficiency is often presented as a \textit{survey-averaged completeness} that can be used to make exoplanet population inferences across the stellar sample. Because planet occurrence rates do vary with stellar properties \citep{Howard2012,Mulders2015}, a description of the stellar sample is needed to place these inferences in context with those of other surveys that target different groups of stars, such as microlensing surveys \citep[e.g.][]{Suzuki2016,pascucci2018universal}. A description of the stellar sample is therefore also included in the Tier I data products.

% bold-face this line as the "one important thing"?
The Tier II data products include the per-star detection efficiency. Providing this data could enable additional science cases such as investigating stellar dependencies in the planet population, and is especially useful for synergies with other detection methods.

\subsection{Tier I Data Products}

\begin{enumerate}
    \item {\bf Stellar sample properties}
    \begin{enumerate}
        \item Description of stellar sample that was searched. This includes target selection criteria and number of stars in the survey.
        \item Provenance of stellar parameters, in case multiple catalogues are available.
        \item Description of the properties of stars that were removed from the parent sample (e.g. possible evolved star or binary, lack of high quality data, etc.)
    \end{enumerate}

    \item {\bf Survey parameters}
    \begin{enumerate}        
        \item Description of criteria used for signal detection and planet candidate vetting (SNR threshold, minimum number of transits required, etc.)
        \item Survey-level summary of observational coverage such as a window function, or duty cycle and time coverage
        \item Description of treatment of planet multiplicity. Procedure to search the lightcurve  for additional transit signals.
        \item Detection and vetting efficiency of survey; either estimated or based on injection-recovery tests.
        \item Treatment for stellar binarity and flux dilution/contamination from nearby or background stars.
    \end{enumerate}

    \item {\bf Planet catalog properties}
    \begin{enumerate}
        \item Properties of each host star: mass, radius, effective temperature, and surface gravity. If available, the limb darkening coefficients used for deriving the planet to star radii from the transit depth. 
        \item Observed properties of each planet candidate: orbital period, transit depth (or planet-to-star radius ratio), and transit duration. If available, estimates of the impact parameter or orbital eccentricity. % (or a/Rs)
        \item Sample reliability (measured or estimated, or assumptions made), including both the astrophysical false positive probability and the instrumental false alarm rate.
    \end{enumerate}
\end{enumerate}

\subsection{Tier II Data Products}

\begin{enumerate}
    \item {\bf Stellar sample properties}
    \begin{enumerate}
        \item Full list of surveyed stars with fundamental properties (identifier, radius, mass, effective temperature, limb darkening coefficients, age estimates, etc.) and their provenance.
        \item  Stellar properties used in defining the survey sample, if those are updated during or after the survey.
        %If updated during after/survey, reflecting current best knowledge
    \end{enumerate}

    \item {\bf Survey properties}
    \begin{enumerate}
        \item Per-target detection efficiency, either estimated from a detection model or as empirically measured with injection-recovery tests. In terms of orbital period and planet-to star radius ratio (or depth). Alternatively, the per-target detection efficiency can also be expressed in terms of the measured signal-to-noise (or MES) and the number of transits observed.
        \item Per-target observing window length including gaps, or 1-sigma depth functions
        \item The detection efficiency of additional planets in the system.
    \end{enumerate}
    
    \item {\bf Planet catalog properties}
    \begin{enumerate}
        \item Simulated data sets.
        Inputs and results of a large suite of simulated data sets based on the pipeline used to generate the planet catalog. Simulated data sets might be based on an astrophysical model plus a model for the noise, or make use of alternative strategies such as scrambling or inverting the data.
    \end{enumerate}
\end{enumerate}

\section{Radial Velocities}
Doppler surveys operate by measuring the change in a star's velocity between numerous, high resolution, spectroscopic observations, and can measure the orbital period, eccentricity, and minimum mass of planets orbiting the target star. Because radial velocity (hereafter, RV) detections do not require an exoplanet to transit its host star, this method is sensitive to a wider range of orbital configurations and to longer period planets than the transit surveys described above. A planet's RV amplitude is a function of its orbital period, mass, eccentricity and inclination -- as such massive, short period, planets seen edge-on are generally easier to detect while low mass, long period planets are more challenging.

Current RV exoplanet surveys tend to target older stars when performing uninformed planet searches, as stars older than about 1\,Gyr are much more RV-quiet than their younger counterparts. RV sensitivity to a planet is determined by the number of RV measurements, the single point measurement precision of these measurements, the ratio of the observational baseline to the planet's orbital period, the cadence and windowing of the observations, instrumental precision and systematics, and stellar variability created by phenomena such as granulation, magnetic activity and spot rotation. Quantifying the Doppler sensitivity for a particular star given an existing RV data set is generally accomplished via the use of injection and recovery analyses.\footnote{For a detailed discussion of such studies and links to an open source python package for running this type of analysis see, e.g., \citet{Rosenthal2021}.} 

\subsection{Tier I Data Products}

\begin{enumerate}
    \item {\bf Stellar sample properties}
    \begin{enumerate}
        \item Catalog of target stars including information on their spectral types, magnitudes, and masses
        \item Total time span of observations for each target star, including start and end dates
        \item Number of RV epochs for each target star
        \item Median RV precision for each star's RV data set
    \end{enumerate}

    \item {\bf Survey parameters}
    \begin{enumerate}        
        \item Time baseline of survey, start and end
        \item Average number of observations per survey target star
        \item Estimated survey average detection efficiency as a function of RV semi-amplitude ($K$) and orbital period derived from injection recovery tests\footnote{Given the way most existing RV surveys are carried out, there is no guarantee that the Doppler sensitivity will be the same even for stars of the same spectral type and magnitude. As RV observations require targeting each star individually, stars generally end up with varying numbers of observations and levels of RV precision depending on the science goals of the survey.}
    \end{enumerate}

    \item {\bf Planet catalog properties}
    \begin{enumerate}
        %\item Properties of each host star: mass, effective temperature, etc. %isn't this already addressed in the Stellar sample section?
        \item Properties of each planet candidate: orbital period, RV semi-amplitude, orbital eeccentricity and longitude of periastron $\omega$ (if used), time of conjunction of periastron (if used), $m$sin$i$, and some quantification of the signal strength (FAP, $\Delta$BIC, etc.)
    \end{enumerate}
\end{enumerate}

\subsection{Tier II Data Products}

\begin{enumerate}
    \item {\bf Stellar sample properties}
    \begin{enumerate}
        \item Target selection function (for instance, any cuts applied on rotational velocity, apparent magnitude, spectral type, stellar activity, stellar multiplicity, etc)
        \item Properties of each host star: apparent magnitude, spectral type, effective temperature, coordinates, mass, radius, distance; in addition, any of the following if known: activity indicators, rotation period, age, metallicity, multiplicity, rotational velocity
        \item Median single measurement precision for each target star
        \item \textbf{RV time-series for each target star, including BJD times of observation, RV, RV uncertainty (due to photon + instrument uncertainties), and SNR in a representative echelle order.} Or if data is still proprietary then the above list without the actual RVs, but still including the BJD times of observation, RV uncertainty, and SNR.
        \item Details of stellar activity related modeling, if performed in planet detection:
        \begin{itemize}
            \item Activity or ancillary spectroscopic indicator time-series:
            \begin{itemize}
                \item Canonical activity indices such as S-index, H-alpha, etc, if used
                \item CCF bisector time-series, as a whole, and/or line-by-line depending on RV extraction methodology, if used
                \item Line-by-line flux-depth time-series if utilized
                \item Order-by-order chromatic RV measurements if utilized
                \item Other activity indicator time-series
                \end{itemize}
            \item Specification of any kernel used for activity temporal modeling, including Gaussian Process (GP) hyperperameters such as spot lifetime, rotation period, GP amplitude, smoothness, etc.
            \item Details on any photometry or other time series data used to estimate the star's rotation period, oscillation/granulation strength, etc, if these parameters were used in the planet search process.
        \end{itemize}
    \end{enumerate}

    \item {\bf Survey parameters}
    \begin{enumerate}        
        \item Details of methodology, priors for planet fit parameters, etc.
        \item Telescope(s) and spectrograph(s) used, including spectrograph configuration(s) if applicable
        \item Any specialized cadence or observing constraints (e.g. bright time only)
        \item Per-star 50\% detection probability as a function of msini and orbital period
        \item Recovery results for each period and $K$ combination injected into the RV data, in a tabular format
        \item Notes on any targets that were added or dropped during the survey, or targets for which the priority of observation planning was changed mid-survey due to evidence of a planet candidate, etc. In the ideal case, this would take the form of an additional column in the RV time series that would denotes why each observation was triggered (e.g., part of a given survey’s algorithmic observations, part of a small proposal to get extra observations for a few targets, because the observing team gave it a priority increase, etc.).

    \end{enumerate}

    \item {\bf Planet catalog properties}
    \begin{enumerate}
        \item Detailed planet orbital information: period, semi-amplitude, eccentricity and longitude of periastron $\omega$ (if used), mean anomaly, and some quantification of the signal strength (e.g. FAP or $\Delta$BIC), etc.
        \item Posterior samples, annotated with log prior and log likelihoods separately, so that other authors can reweight the sample to account for different priors. Particularly, for low SNR detections where uncertainties can be large and non-Gaussian.  

        \item Check on whether the signal's period also appears in the star's activity time series
        \item Check on whether the signal's period could be due to survey design (e.g. aliasing, window function, etc.)
    \end{enumerate}
\end{enumerate}

\section{Direct Imaging}

Direct imaging requires overcoming the extreme contrast ratios and small angular separation between planet and host star.  Adavances in adaptive optics on large ground-based telescopes and coronagraphy have made way for the detection of exoplanets at small angular separation (e.g. \citealt{macintosh2018,beuzit2019}). Most direct imaging exoplanet surveys target young planets in the infrared, as young planets are hotter and brighter, and present a more favorable contrast ratio to their host star \citep{bowler2016}.  More advanced imaging instrumentation on large telescopes, coupled with large target lists of young, nearby stars have enabled direct measurements of the demographics of young, giant planets ($\gtrsim 1$M$_{Jup}$) at wider separations ($\sim$10-100 au) around nearby stars (e.g. \citealt{Nielsen2019,vigan2021}).
%\ilaria{MagAO is using optical wavelengths but that's because they target young accreting planets, perhaps this should be mentioned or maybe start with "Most current direct imaging..."}  -- Good point, updated text accordingly (Eric)

Imaging observations of a given star are characterized by a contrast curve, the minimum flux of a planet that is detectable as a function of projected separation from the star. Demographic studies from imaging data use these contrast curves and target properties to translate the observing sensitivity into a completeness to planets.  Theoretical models (e.g. \citealt{baraffe2003,marley2021}) provide spectra of exoplanets as a function of planet mass and system age. Then, given the age and apparent magnitude of each target star, these theoretical models are used to convert the contrast curve from flux ratio units (in a given passband) into planet mass. Typically, projected separation is also converted to orbital semi-major axis by marginalizing over possible orbital parameters. Given the limited number of young, nearby stars in the sky (for example, those in nearby moving groups, \citealt{gagne2018}), different demographic studies often have overlapping target lists.
%\rachel{(Rachel: I think this paragraph needs a little more to be similar in length to the transit and RV sections. Also, add citations.)} \ilaria{I agree and it would be good to briefly mention how they convert a planet luminosity into a planet mass} -- Sounds good, added that in (Eric)

\subsection{Tier I Data Products}

\begin{enumerate}
    \item {\bf Stellar sample properties}
    \begin{enumerate}
        \item Description of stellar sample that was searched, including target selection criteria and number of stars in survey
        \item Properties of each target star: age, distance, magnitude in the observing band
        \item Provenance of stellar parameters
    \end{enumerate}
    \item {\bf Survey properties}
    \begin{enumerate}
        \item Description of the criteria used for detection
        \item Description of image processing method, and characterization of any bias introduced
        \item Contrast curves (in units of delta-magnitude as a function of projected separation in arcseconds) for each observation of each target star (with and without planets). If azimuthally-collapsed 1D contrast curves are produced, the 2D images should be preserved
        \item Noise estimation method
    \end{enumerate}
\item {\bf Planet catalog}
    \begin{enumerate}
%    \item Host star properties (age, distance, magnitude in the observing band)
        \item Observed properties of each companion: delta-magnitude, epoch of detection, projected separation at detection, and position angle
        \item Theoretical model used to convert planet luminosity to planet mass
        \item False alarm probability or other method/s of characterizing sensitivity
    \end{enumerate}
\end{enumerate}

\subsection{Tier II Data Products}

\begin{enumerate}
    \item {\bf Stellar sample properties}
    \begin{enumerate}
        \item Probability distributions for stellar age and mass for each star, if generated.
    \end{enumerate}
    \item {\bf Survey properties}
    \begin{enumerate}
        \item 2D plots (and underlying data) showing fractional completeness to planets around each star as a function of semi-major axis and planet mass (depth of search plots, or "tongue plots")
        \item Raw and processed images after completion of a given demographics study to enable reanalysis with new techniques.
    \end{enumerate}
    \end{enumerate}

\section{Microlensing}

The microlensing technique enables the measurement of the planet-host mass ratio, $q$, and the projected orbital separation of a planet relative to the size of the Einstein ring radius of the host star, $s=a_\perp/R_{\rm E}$  \citep{Gaudi2012}. Microlensing events are inherently rare, requiring the near perfect alignment of a lensing system with a background source star. As such, microlensing surveys monitor the Galactic Bulge where the event rate peaks with a large number of potential sources and planetary lens systems.
The primary advantage of the microlensing method is its sensitivity to low-mass planets \citep{bennett96} in relatively wide orbits. The microlensing method is most sensitive to planets orbiting at separations within a factor of two of the Einstein radius, which is typically 2-$3\,$AU \citep{gould-loeb92}. Microlensing's unique sensitivity to low-mass planets in wider orbits was the rationale for the selection of an exoplanet microlensing survey as one of the main goals of the top ranked \textit{Nancy Grace Roman Space Telescope} (formerly \textit{WFIRST}) by the Astro2010 decadal survey.

With one notable exception \citep{Nucita_Kojima1}, all planetary microlensing events have been detected toward the Galactic bulge, including the discoveries by the Optical Gravitational Lens Experiment (OGLE), the Microlensing Observations in Astrophysics group (MOA), and the Korean Microlensing Telescope Network (KMTNet). The \textit{Roman} Galactic Bulge Time Domain Survey\citep{Penny2019} will also monitor the central Galactic bulge in order to carry out its exoplanet microlensing survey. 
As the distribution of lens systems studied by these surveys extends from near the Earth to the Galactic center, microlensing can detect planets throughout the Galaxy. The parameters measured directly from virtually all planetary microlensing events are $q$ and $s$, the mass ratio and separation in Einstein radius units. Some have argued that the mass ratio is more fundamental than the planet mass \citep{pascucci18}, and for microlensing events it is more sensible to characterize planetary systems with planetary mass ratios and host star masses, instead of the masses of the host and the planets. The determination of the host star masses and their distance from us does not automatically come from the light curve models and relies upon additional observations and/or the measurement of subtle light curve features as we discuss in Section~\ref{ML-mass-dist}.

The characteristic timescale of a microlensing event, $t_{\rm E}$, for a stellar lens is typically tens of days, but ranges from four to $>100\,$days for known planetary microlensing events. Light curve perturbations due to planets have a duration that can be less than a few hours.. A balance must be made between observation cadence and survey area to detect as many events as possible but maintain a cadence to characterize planetary anomalies. Microlensing exoplanet searches often employ auxiliary observations targeted at events that showed early potential of exhibiting a planetary signal. If these auxiliary observations are scheduled based on a possible planetary signal, this has the potential to bias the results if the triggering of these auxiliary observations is not properly accounted for in the statistical analyses. However, the selection of high magnification events \citep{gould10} or bright source stars \citep{gaudi02,cassan12} because of their high sensitivity to exoplanet signals does not cause such a bias. Since follow-up observers are strongly tempted to observe events with possible planetary signals in progress, it is helpful to define a protocol to ensure a well defined statistical sample in advance \citep{Yee2015}.

High cadence microlensing surveys with wide field telescopes \citep{Suzuki2016,Poleski21} can survey many more microlensing events than follow-up programs and therefore have a higher planetary signal detection rate. So, we expect that most future exoplanet microlensing demographics papers will be based on high cadence surveys. Nevertheless, data taken in response to the identification of a possible planetary signal can still be used to more precisely determine the lens system properties, as long as the originally scheduled survey observations are sufficient to establish that the possible planetary signal is above the detection threshold \citep{Suzuki2016}.

% McDonald2021 is not a demographics paper.
% There is no bias from auxiliary observations if they are scheduled based on high planet detection sensitivity instead of
% the detection of a possible planetary signal.

The exoplanet microlensing demographics papers to date \citep{gaudi02,sumi10,gould10,cassan12,Suzuki2016,Poleski21} have all used the planet-star mass ratio, $q$, and the separation on the plane of the sky in Einstein radius units, $s$, to characterize the planetary systems. This is done because, this information comes directly from the light curve model for virtually all events, although there are some events with degenerate light curve models, particularly the close-wide degeneracy that yields nearly identical light curves for most high magnification events with the substitution $s \leftrightarrow 1/s$.

One of the main challenges for the microlensing method is determine the masses and distances of the planetary systems discovered with this method. This would enable a more direct comparison to the demographics from radial velocity and direct imaging surveys, although some progress can be made without mass and distance measurements \citep[e.g.,][]{Clanton2016}. The next major advance in exoplanet microlensing is the \textit{Nancy Grace Roman Space Telescope} (\textit{Roman}) Galactic Bulge Time Domain Survey \citep{Penny2019}. \textit{Roman} will produce a statistical sample of planets comparable to \textit{Kepler}'s results from $\lesssim1$ au, but in a complementary region of parameter space from 1 to $\gtrsim$10 au. In addition to a much larger sample of microlens planetary systems, \textit{Roman} will determine the masses and distances for more than half the planetary systems that it discovers, using the methods described in the next section.

\subsection{Microlens Mass and Distance Measurement Methods}
\label{ML-mass-dist}

\begin{figure}[h]
 \begin{minipage}[t]{0.5\textwidth}
  \includegraphics[width=\textwidth]{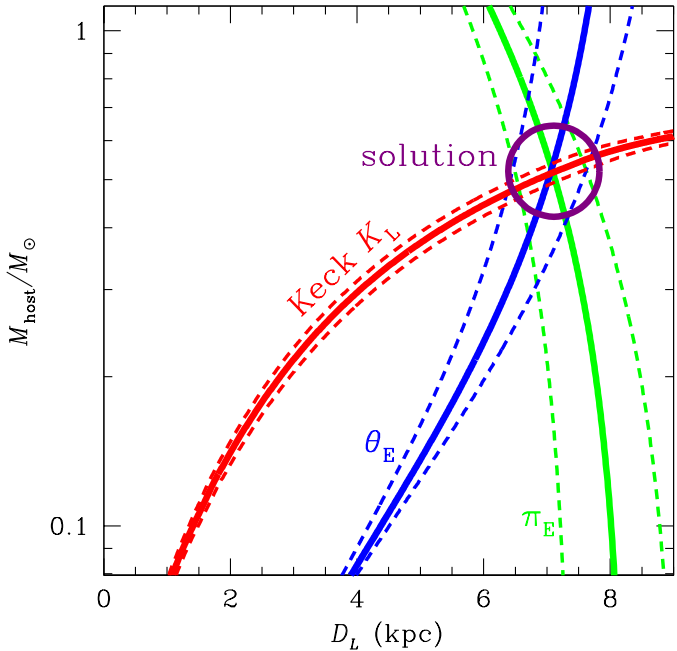}
 \end{minipage}\hfill
 \begin{minipage}[b]{0.45\textwidth}
\vspace{-0.4cm}
\caption{\sl \small 
Three mass-distance relations are plotted for planetary microlensing event
MOA-2009-BLG-319. The constraints from the measured angular Einstein radius, $\theta_E$, are
shown in blue; the Keck $K$-band lens brightness measurement, $K_L$ is shown in red, and the 
microlensing parallax, $\pi_E$, constraint is shown in green (although $\pi_E$ was not actually measured
for this event). The dashed lines are 1-$\sigma$ error bars.
The intersection of these curves yields the lens system mass and distance, and the planet-star
mass ratio determined from the microlensing light curve gives the individual star and planet masses.}
\label{fig-mass-dist}
 \end{minipage}
\end{figure}

There are three methods that yield mass-distance relations for microlensing events: measurement of the angular Einstein radius, $\theta_{\rm E}$, measurement of the microlensing parallax parameter, $\pi_{\rm E}$, and the brightness of the source star. Figure~\ref{fig-mass-dist} shows these mass distance relations for planetary microlensing event MOA-2009-BLG-319 (although $\pi_{\rm E}$ was not actually measured for this event). Any two of these measurements can yield a mass and distance solution, and all three of them give redundant constraints that can be used to check for systematic errors.

The angular Einstein radius can be determined from finite source effects that are measured for most planetary microlensing events, particularly those with low-mass planets via $\theta_{\rm E} = \theta_* t_{\rm E}/t_*$, where $\theta_*$ is the angular radius of the source star, which is determined from its extinction corrected magnitude and color. The microlensing parallax parameter, $\pi_{\rm E}$, is the inverse of the length of the Einstein radius, projected to the position of the observer, and it can be measured when the signature of the Earth's orbit is seen in light curve \citep{muraki11} or from observations from a telescope in a Heliocentric orbit \citep{udalski-ogle124}. For events with caustic crossings or planetary features, observations from Earth and L2 can also measure $\pi_{\rm E}$ \citep{gaia16aye}. However, both ground-based and space-based $\pi_{\rm E}$ measurements are susceptible to systematic photometry errors \citep{koshimoto20_spit_err,koshimoto21,yee21}, so some care is required when interpreting the $\pi_{\rm E}$ values from light curve models.

The final method to determine a mass-distance relation is to measure the brightness of the planetary host star (which is also the lens star). For ground-based surveys, this can be accomplished with high angular resolution follow-up observations using the \textit{Hubble Space Telescope} or a large ground-based telescope with a laser guide star adaptive optics system \citep[e.g.,][]{bennett15,batista15}.  These high angular resolution observations are needed to resolve the lens star separating from the source star in order to avoid confusion with other nearby stars other than the lens (and planetary host) \citep{aparna17,koshimoto20hires}. For a space-based exoplanet microlensing survey, like the one planned for \textit{Roman}, the survey data have high enough angular resolution to detect the lens stars and measure the lens-source relative proper motion, $\murelbold$ \citep{bennett-gest,bennett07}. The measurement of $\murelbold$ is also useful for confirming the lens star identification since the length of the $\murelbold$ vector can be determined from light curve parameters, $\mu_{\rm rel} = \theta_*/t_*$. In addition the $\murelbold$ and $\piEbold$ vectors have the same direction, so a measurement of $\murelbold$ can be used to test for systematic errors in $\piEbold$ measurements or to provide precise $\pi_{\rm E}$ measurements when the light curve only imposes a strong constraint on one component of $\piEbold$ \citep{aparna18,bennett20}.

We should also note that these three mass-distance relations are also useful for demographics studies even in cases where only one of the mass-distance constraints is measured precisely \citep{koshimoto21}. Of course, a single mass-distance constraint will produce a mass-distance degeneracy in the interpretation. However, a typical microlensing survey will have some events with mass measurements, and some with a combination of a single mass-distance relation with upper and/or lower bounds on the other mass-distance relation parameters. For microlensing event MOA-2010-BLG-477, a measurement of $\theta_{\rm E}$, upper and lower limits on $\pi_{\rm E}$, and an upper limit on the host star brightness imply that the host of the gas giant planet is a white dwarf \citep{blackman_mb10477}. %\ilaria{The microlensing section provides a nice and detailed overview of the technique, should the other methods follow a similar approach?}

\subsection{Tier I Data Products}

\begin{enumerate}
    \item {\bf Stellar sample properties}
    \begin{enumerate}
        \item List of all microlensing events in the sample including coordinates and measured $t_{\rm E}$, $t_0$, and $u_0$ values. When $t_*$ is measured, it should be reported, too. For surveys including binary star host systems, the binary microlensing model parameters should also be included.
    \end{enumerate}

    \item {\bf Survey properties}
    \begin{enumerate}
        \item Description of detection thresholds for events included in sample (e.g., $\Delta\chi^2$ cuts)
        \item Average detection efficiencies across the survey as a function of $q$, $s$, and $t_{\rm E}$, or 
        \item Method for estimating source radius crossing time, $t_*$, or equivalently, the source size in Einstein radius
                 units ($\rho = t_*/t_E$) for events without planetary signals or finite source effects. This is used for detection
                 efficiency calculations.
    \end{enumerate}

    \item {\bf Planet catalog properties}
    \begin{enumerate}
        \item Observed properties of each microlensing event: mass ratio ($q$), projected separation ($s$, in Einstein radius units), angle between source trajectory and lens axis, $\alpha$, and source magnitude in one or more passbands, with error bars. For events with finite source effects (most, but not all planetary events), also include the source radius crossing time, $t_*$. For events with detected multiple planet systems, the additional parameters needed to describe the event should also be included.
        \item Mass-distance relation properties of each host star
            \begin{itemize}
                \item microlensing parallax, $\piEbold$, a two-dimensional vector, with error bars
                \item angular Einstein radius, $\theta_{\rm E}$, usually calculated from finite source effects
                \item host star brightness or limits in one or more passbands
                \item relative lens-source proper motion, $\murelbold$, a two-dimensional vector, as measured from high angular resolution observations before or after the event
            \end{itemize}
        \item For events with degenerate models, include $t_{\rm E}$, $q$, $s$, $\alpha$, and $t_*$ for all degenerate models with error bars; also include a weighting for each model, with the calculation method used to determine the weighting. 
    \end{enumerate}
\end{enumerate}

\subsection{Tier II Data Products}

\begin{enumerate}
    \item {\bf Stellar sample properties}
        \begin{enumerate}
        \item Distribution of host star and stellar remnant masses surveyed for planets and the Galactic model used to calculate this distribution.
    \end{enumerate}
    \item {\bf Survey properties}
    \begin{enumerate}
        \item {\bf Planet detection efficiencies as a function of $s$ and $q$ for each microlensing event in the sample.} Ideally, this would extend upward through mass the mass ratios corresponding the brown dwarf and stellar mass secondaries, although most analyses to date have not done this. This enables the straight forward combination of multiple microlensing demographic analyses, and it is probably the most useful of the exoplanet microlensing demographics data products.
%        \item \gijs{Galactic stellar population model used to derive planet mass and/or semi-major axis. Description of the lens luminosity/distance bias in survey.}
    \end{enumerate} 
\end{enumerate}

\section{Astrometry}

Astrometry involves measuring the positions of celestial bodies. As such, it provides two-dimensional information---typically in the right ascension and declination axes---about objects that give information about trajectories on the sky (proper motions), as well as distances (parallaxes). In addition, astrometry yields information on the motions of bodies in gravitationally bound systems involving multiple objects, such as binary asteroids, moons orbiting planets, trans-Neptunian objects, and stars. Over the past few decades, astrometric capabilities have advanced to the point where detections of exoplanets orbiting other stars are within reach \citep{Lurie2014,Boss2017,Kervella2022}. This section describes data requirements for exoplanets orbiting other stars that are detected via astrometry.

%\rachel{(Rachel: Is there any point to mentioning the planet list that Gaia promised?}
%\eliot{Eliot: I think this could be done if we used it as a tool to highlight how the technique works in practice and why it's difficult to detect exoplanets with astrometry.)} \ilaria{The microlensing section mentions specific surveys/missions, I think it would be valuable to briefly discuss Gaia here}

The resulting astrometric signal maps the position of the combined light of the target system, which means in the case of an exoplanetary system the observed astrometric displacements represent only those of the host star. However, the same observational signal may be mimicked by a binary star in which the components have nearly equal fluxes. In that case, the system's observed center of light is only slightly displaced from the system's barycenter. Potential detections must therefore be examined with high resolution imaging techniques to rule out the possibility that a non-planetary mass luminous companion is causing the astrometric orbital signature.

The items below describe the minimum data products that should be provided for surveys of stars for exoplanet companions via astrometry, followed by additional recommended data products.

\subsection{Tier I Data Products}

\begin{enumerate}
    \item {\bf Stellar sample properties}
    \begin{enumerate}
        \item Number of stars in survey
        \item Description of the stellar sample that was searched, including target selection criteria (RA/Dec, distance (preferably trigonometric parallax), absolute magnitude, photometric color)
        \item Stellar masses and method by which those were determined. This property should already be known for at least every planet detection, as the stellar mass is necessary to distinguish if a signal is from a planetary companion or a stellar companion.
    \end{enumerate}

    \item {\bf Survey properties}
    
    \begin{enumerate}
        \item Survey facility information (telescope, including aperture; instrument, including pixel scale and filter(s) used)
        \item Representative value of the number of observing epochs (nights/months/years) typical per target
        \item For each target, the five-parameter astrometric solutions (position = RA+Dec, parallax, and proper motions in RA+Dec with uncertainties)
        %
        % #2 most important thing = total survey duration
        \item {\bf Total survey duration, including start and end dates and representative value of the total years of observational coverage per target} 
        %        
        % #1 most important thing = typical sensitivity per target
        \item {\bf Typical sensitivity per target in milliarcseconds at orbital periods of up to $N$ years. This property essentially summarizes the implications of the survey properties listed above. Most surveys will be doing this already, on at least a target-by-target basis, as part of their search for potential planetary signals.}
        \item For targets without detections, typical sensitivity in Jupiter masses at orbital periods up to $N$ years (assuming stellar mass and various stated orbital parameters). This property may take some additional analysis to produce, but it is necessary for any comparisons with other results to be meaningful. It also indicates if the validity of these non-detections will need to be reassessed when any of the dependent properties (such as stellar mass, stellar multiplicity, or possible planet orbits) are later updated. 
    \end{enumerate}
    
    In particular, the above properties (a) though (d) should be well known for any survey target without additional effort.

    \item {\bf Planet catalog properties}
    \begin{enumerate}
        \item For each detection, the photocentric orbit parameters and their uncertainties, with a thorough description of the astrometric model, including how the following seven parameters are determined simultaneously with the five basic astrometry parameters:
        \begin{itemize}
            \item {\it P} = orbital period
            \item {\it a} = semi-major axis
            \item {\it e} = eccentricity
            \item {\it i} = inclination
            \item $\omega$ = argument of periastron
            \item $\Omega$ = longitude of ascending node
            \item $T_o$ = epoch of periastron
            \item equinox of the grid used to set angles
        \end{itemize}
        These orbital elements should also be reported in the form of the Thiele-Innes constants (A, B, F, and G).
        
        These parameters should already be produced for every planet detection as part of the analysis to determine if the signal is from a planetary companion or a stellar companion.
        \item For each detection, a figure showing the perturbations on each axis (RA vs.~time and Dec vs.~time) with respect to the astrometric model fit. The data behind these figures should also be made available in machine-readable format.
        \item Some measure of the probability that a given signal is not explained by a 5-parameter single-object astrometric model.  This is more appropriate for a large survey result, rather than a report of a planet (or planets) orbiting a given star.
        \item Contrast curves from high-resolution imaging of the targets to eliminate the possibility that the astrometric signal is due to a luminous stellar or brown dwarf companion, rather than a planet. This data product should already be produced for every planet detection as part of that analysis.
    \end{enumerate}
\end{enumerate}

\subsection{Tier II Data Products}

\begin{enumerate}
    \item {\bf Stellar sample properties}
    For each target star: 
    \begin{enumerate}
        \item Stellar luminosity and determination method
        \item Stellar temperature and determination method
    \end{enumerate}

    \item {\bf Survey properties}
    \begin{enumerate}
        \item Total duration of observations on each individual target, including start and end dates and the total years of observational coverage
        \item Number of observing epochs (nights/months/years) on each target
        \item For each target, sensitivity in milliarcseconds at orbital periods of 1--$N$ years, with values provided at key orbital periods within that range. This property will require some analysis of the residuals to the proper motion and parallax fits, but many surveys will likely be doing this already as part of their search for potential planetary signals.  
        \item For each target without detections, sensitivity in Jupiter masses at orbital periods of 1--$N$ years (assuming stellar mass and various stated orbital parameters), with values provided at key orbital periods within that range. These curves should thus capture the mass limits of companions ruled out by the astrometric time series. This information can be concisely presented for each target in a single figure, for example as in Figure 4 of \citet{Lurie2014}. 
        
        Each such curve may take some additional analysis to produce, but it will enable meaningful comparisons with results covering different ranges of orbital period space. These curves also allow the validity of these non-detections to be reassessed on a target-by-target basis if any of their dependent properties (such as stellar mass, stellar multiplicity, or possible planet orbits) are later updated. 
        \item A periodogram of the astrometric residuals to the proper motion and parallax fits on each directional axis (RA and Dec). 
        \item Data frames that are flat-fielded and bias-subtracted, as well as plate constants specific to those frames.
        \item Centroid positions in RA and Dec at each epoch for the target star and each reference star used in the analysis.
    \end{enumerate}
% 4:30 p.m. on Weds Dec 8: Eliot paused while editing here...
    \item {\bf Planet catalog properties}
    \begin{enumerate}
        \item For each detection, a figure showing the photocentric orbit in the plane of the sky (RA vs.~Dec) with proper motion and parallax removed. This product will be helpful for comparing other results to this detection, in particular for audiences unaccustomed to astrometry data. 
        \item For non-detections, a figure showing the photocenter residuals after solving for parallax and proper motion on each axis (RA vs.~time and Dec vs.~time)
    \end{enumerate}

\end{enumerate}

\section{Data archiving}

The sections above have outlined the data products that the ExoPAG SIG on Exoplanet Demographics finds would be valuable when published alongside exoplanet survey data, in order to allow for the usage of those survey data by the wider exoplanet demographics community. However, one topic that we did not cover was the logistics of archiving these data such that they are reliably and permanently accessible. We recognize that some of the data products are non-trivial in size, and/or do not naturally conform to the data standards typically accommodated by journals. We also want to encourage authors to make their simulated data sets available where possible, which will incur additional overheads. 

There are several archiving options that are already available to the community. The Direct Imaging Virtual Archive (DIVA)\footnote{Direct Imaging Virtual Archive: \url{http://cesam.lam.fr/diva/}} houses high-level science products (HSLPs) across multiple direct imaging detection surveys, providing an example of a repository for the kinds of data that are needed to perform demographics studies across surveys. In particular, DIVA encourages the use of a consistent format, HCI-FITS, for the uploading and storage of HLSPs, ``to simplify HLSP exchange in the HCI community by adopting a standard output format for all the usual data products". DIVA currently hosts nearly 20\,GB of data across 16 direct imaging surveys, and invite other surveys to contribute data.

The NASA Exoplanet Archive hosts contributed datasets from the community\footnote{NASA Exoplanet Archive: Contributed Data Sets: \url{https://exoplanetarchive.ipac.caltech.edu/docs/contributed_data.html}} and a substantial amount of ancillary survey data and HLSPs from the NASA \kepler mission, including detailed characterizations of the survey detection efficiency and reliability of the types described here\footnote{Kepler Completeness and Reliability Products: \url{https://exoplanetarchive.ipac.caltech.edu/docs/Kepler_completeness_reliability.html}}\footnote{Kepler Simulated Data Products: \url{https://exoplanetarchive.ipac.caltech.edu/docs/KeplerSimulated.html}}. They also host a large number of transiting and radial velocity survey data sets, and are open to hosting additional data sets by request. 

The VizieR Information Service\footnote{ViZieR: \url{https://vizier.u-strasbg.fr/viz-bin/VizieR}}, hosted by the Strasburg astronomical Data Center (CDS), provides a standardized description of astronomical catalogues and access to the data therein. Large tables of data, such as stellar and planet catalog parameters, and injection/recovery results, are able to be hosted and accessed in a variety of ways.

The Mikulski Archive for Space Telescopes (MAST) hosts a wide-variety of community-created HLSPs\footnote{MAST HLSPs: \url{https://archive.stsci.edu/hlsp/}}. These HLSPs contain observations, catalogs, models or simulations that complement, or are derived from, MAST-supported missions. These missions include \emph{Hubble, Webb, Kepler, TESS} and \emph{Roman}. 

In addition to the above options, there are also typically hosting options available at the authors' home institutions, however there can be questions about the long-term storage and accessibility of these ad hoc sites. We encourage authors to find a permanent archival home for their data products, to ensure their ongoing accessibility and usage.


\clearpage

\bibliography{sample631}{}
\bibliographystyle{aasjournal}



\begin{acknowledgements}

The authors would like to gratefully acknowledge those members of the community who provided feedback on initial drafts of this report. These include Eric Ford, Timothy Brandt, Andreas Quirrenbach, Jonathan Zink, Michelle Kunimoto, and Eric Mamajek. Thank you for your careful attention and helpful suggestions!

\end{acknowledgements}

\end{document}